\documentclass{jltp}

\usepackage{graphicx} % uncomment this line to include the graphicx package

\title{Antiferromagnetic order and dielectric gap within the vortex core of
antiferromagnetic superconductor}

\author{V. V. Garkusha and V. N. Krivoruchko$^*$}

\address{Donetsk National Universty, Str.\\ Universitetskaya 24, 83055
Donetsk, Ukraine\\
$^*$Donetsk Physics \& Technology Institute NAS of Ukraine, \\ Str. R. Luxemburg 72,
83114 Donetsk, Ukraine}

\runninghead{V. V. Garkusha and V. N. Krivoruchko}{ Vortex core of
antiferromagnetic superconductor}

\begin{document}

\maketitle

\begin{abstract}
The structure of a superconducting vortex has been studied theoretically for a
dirty antiferromagnetic superconductor (AFSC), modelling an AFSC as a doped
semi-metal with s-wave superconducting pairing and antiferromagnetic
(dielectric) interaction between electrons (holes). It is also supposed that
the quasiparticles dispersion law possesses the property of nesting. The
distribution of the superconducting and magnetic order parameters near the
vortex core is calculated. It is shown that the antiferromagnetic order, been
suppressed at large distances, is restored around the superconducting flux and
the vortex core is in fact insulating and antiferromagnetic, in stark contrast
to the normal metal cores of traditional superconductors. Moreover, our model
calculations predict that as the temperature decreases the flux region of the
superconductivity and antiferromagnetism coexistence increases.

\end{abstract}

\section{INTRODUCTION}

The influence of the antiferromagnetic (AF) correlation in the superconducting
state has been attracting much attention recently. In particular, the
interrelationship between magnetism and superconductivity (SC) have been
actively discussed in connection with the problem of the symmetry type of the
order parameter in high-$T_{C}$ systems (see, e.g.,$^{1}$). It is natural also
to ask if the close proximity between antiferromagnetic and
superconducting\ phases could have any macroscopic manifestations. At the
macroscopic level, the external magnetic field penetrates type-II
superconductors via normal state metallic inclusions, or vortices, and can reveal the nature of the system's
ground state. So that, by weakening the superconductivity of high-$T_{C}$ due
to magnetic field, the interplay between antiferromagnetic and
superconducting\ ordering can be explored. There are predictions that vortices
in high-$T_{C}$ cuprates differ from those in conventional superconductors.
E.g., Zhang$^{2}$ has developed a unified theory where antiferromagnetism and
$d$-wave\ SC in high-$T_{C}$ superconductors has been constructed based on
$SO(5)$ symmetry. In this theory the AF and SC order parameters are unified
into a five dimensional vector (a superspin). The $SO(5)$ theory predicts the
existence of superconducting vortices with antiferromagnetic cores as a
spatial class of topological solutions.$^{3}$ Recent magnetic neutron
diffraction data,$^{4}$ scanning tunneling microscopy,$^{5}$ muon spin
rotation$^{6}$ and nuclear magnetic resonance$^{7}$ measurements on some
high-$T_{C}$ SC have revealed that the applied field which imposes the vortex
lattice, also induces antiferromagnetic electronic correlations. The new
finding is that for optimally doped samples (superconducting material shows no
static magnetism in either zero or weak magnetic field) the vortex state has
much stronger tendencies towards magnetic order than either the normal or
superconducting state. Since magnetism extend beyond the vortex core into the
superconducting regions of the material, these result means that
antiferromagnetic correlation and superconductivity coexist throughout the
bulk of the material, and applied field has only effect of recovering some of
the magnetic correlation.$^{3-7}$ One view that this means that the band
quasiparticle picture cannot account for the AF order of the vortices and the
experimental facts provide good support of the idea of the existence of
$SO(5)$ symmetry in high-$T_{C}$ cuprates. (see, e.g.,$^{3}$ and references therein).

Recently it was discovered that competition between superconductivity and
antiferromagnetism also occur in the new classes of quaternary intermatallic
superconductors such as the the borocarbide series $RNi_{2}B_{2}C$ and $RNiBC$
(here $R$ - rare-earth ion) (see, e.g.,$^{8}$ and references therein). For
these new systems, the magnetic energies are dominated by exchange
interactions and comparable to the magnetic superconducting condensation
energies ensure that the interlay between superocnductivity and
antiferromagnetism should be on a comparable energetic footing.

In this report, we study the possibility for antiferromagnetic ordering of the
vortex based on the model where the Hamiltonian consists of two components,
one corresponding to the (Cooper pair) condensation conditions of equal and
opposite momenta and spins and the other to the magnetic interaction condition
of parallel spins and finite momentum transfer. A mean-field approach is taken
to study the magnetic and superconducting structures of a flux vortex in such
AFSC in the so-called "dirty" limit. We show that the external magnetic field
stabilizes the AF order around the vortex core in a region where SC have been
suppressed, i.e., a coexistence phase is induced by the applied field. It was
also obtained that the vortex core is, in fact, insulating in stark contrast
to the standard metal cores of low-temperature superconductors.

\section{The model}

Selecting the Hamiltonian of an AFSC , we will use the isotropic model of a
doped semi-metals with Cooper pairing electrons (holes) within each (two) of
the bands and dielectric (electron-hole) pairing between bands.$^{9}$ In the
self-consistent field approximation we have
\begin{eqnarray}
&&H=\int d\bf{r}\left\{\sum_{\sigma}\left[\psi_{\sigma}^{+}\xi_{1}\psi_{\sigma}+
\varphi_{\sigma}^{+}\xi_{2}\varphi_{\sigma} \right] \right.\nonumber \\
&&\left. + [\Delta_{1S}(\bf{r})\psi_{\uparrow}^{+}\psi_{\downarrow}^{+}%
+\Delta_{2S}(\bf{r})\varphi_{\uparrow}^{+}\varphi_{\downarrow}%
^{+}+h.c.] +\Delta_{A}(\bf{r})\sum_{\sigma}[\psi_{\sigma}^{+}%
\varphi_{\sigma}+h.c.]\right\}
\end{eqnarray}
Here $\psi_{\sigma}^{+}$, $\psi_{\sigma}$ and $\varphi_{\sigma}^{+}$,
$\varphi_{\sigma}$ are the creation and annihilation operators of electrons
and holes in the first and second bands, respectively, whose dispersion law
satisfies the condition of nesting\ $ \xi_{1,2}(\mathbf{p+Q})=\xi
_{\mathbf{p}}$. The fraction of the Fermi surface that is nested, $\nu$ , will
be treated as a variable. As is known, the Fermi surface nesting causes the
transition to spin density wave state.$^{9,10}$ This state can be described by
dielectric order parameter, which characterizes the gap in quasiparticle's
excitations. For the case of an antiferromagnetic (doubled) cell, we take
$\bf{Q}=\bf{K}/2$ , where $\bf{K}$ is the vector of a reciprocal
lattice of the crystal. The operator $\xi_{\bf{P}}$ is understood to be
$\hbar^{2}(\bf{p}-e\bf{A})^{2}/2m-E_{F}$; $\Delta_{1(2)S}(\mathbf{r})$ are
the band superconducting order parameters; $\Delta_{A}(\mathbf{r})$ is the magnetic
order parameter; $E_{F}=\hbar^{2}p_{F}^{2}/2m$ stands for the Fermi energy;
and $\mathbf{A}$ is the vector potential. As one can see, for the model (1)
the SC correlations have the spin-singlet type, while the magnetic ones have
the spin-triplet type.

The system of Gor'kov equations for electrons in the first band will be
distinguished from the standard system for functions $G_{11}(1,2)=-<T_{\tau}%
\psi_{\uparrow}(1)\psi_{\uparrow}^{+}(2)>$ and $F_{11}(1,2)=<T_{\tau}%
\psi_{\downarrow}^{+}(1)\psi_{\uparrow}^{+}(2)>$ by the presence of the additional
function $\Im_{21}(1,2)=-<T_{\tau}\varphi_{\downarrow}(1)\psi_{\uparrow}^{+}(2)>$
which describes the AF correlations of electrons from different bands. 
(Here we used standard notifications, i.e., $T_{\tau}$ is the $\tau$-ordered operator,
$1=(\bf{r}_{1},\tau_{1})$ and so on.) Due to the magnetic interaction between
the electrons of different bands, the Green's functions of the second band
with a reversed sign for the spin appear in set of equations. The complete
system of equations contains 12 equations (see Ref.11 for details). In this
report the case, when total $SC$ order parameter of the system has the s-wave
symmetry, i.e., the case$^{9}$ $\Delta_{1S}=\Delta_{2S}=\Delta_{S}$ will be
studied. We present here the main equations for such a system in so-called
dirty limit.$^{12}$

\section{Quasiclassical equations for dirty AFSC}

There are three characteristic lengths for the system under consideration: the
electrons mean free path $\ell$ , the SC correlation length $\xi_{S}\sim\hbar
v_{F}/\Delta_{S0}$ and the AF correlation length $\xi_{A}\symbol{126}\hbar
v_{F}/\Delta_{A0}$ . Here $\Delta_{S0}$ is the superconducting order parameter
if the antiferromagnetic pairing is absent, while $\Delta_{A0}$ is the
antiferromagnetic one when the superconducting pairing is absent. The $\xi
_{A}$ can be considered as the characteristic length of electron liquid's
magnetic modulation or magnetic stiffness, as well. We will assume that the
metal is \textquotedblright dirty\textquotedblright\ for both orders; i.e.,
the mean free path $\ell$ and the correlation lengths $\xi_{S}$ and $\xi_{A}%
$\ satisfy the condition $\ell\ll(\xi_{S},\xi_{A})$ . Using the standard
procedure, one can obtain$^{11}$ the system of equations, that are the
generalized Usadel equations for the model (1).

It is
convenient$^{11}$ to pass from the band superconducting variables to the total
ones: $\Delta_{S}=\Delta_{1S}+\Delta_{2S}$ , and $F=F_{11}+F_{22}$ ,
$G_{11}=G_{22}=G$ , $\Im_{21}=\Im_{12}=\Im$. There are tree type of Green's
functions: the functions $G\equiv G(\omega,\mathbf{r)}$ and $\Im\equiv
\Im(\omega,\mathbf{r)}$ describe the normal excitations in system of electrons
with antiferromagnetic correlations, while the anomalous Green's function
$F\equiv F(\omega,\mathbf{r)}$ describes the Cooper-pair condensate. These
functions are related by the usual normalization condition
\begin{equation}
G^{2}(\omega,\mathbf{r})+|F(\omega,\mathbf{r})|^{2}+|\Im(\omega,\mathbf{r}%
)|^{2}=1
\end{equation}
Magnetic and superconducting properties of the system are described by
generalized quasiclassical equations that can be written in the form:$^{11}$
\begin{equation}
-\frac{1}{2}D\mathbf{\Pi}(G\mathbf{\Pi}F-F\nabla G)=\frac{\Delta_{S}}{\hbar
}G-\omega F
\end{equation}%
\begin{equation}
D\nabla(G\nabla\Im-\Im\nabla G)=2\omega\Im-2\frac{\Delta_{A}}{\hbar}G
\end{equation}
Where $\mathbf{\Pi}\equiv\nabla+2i\pi\mathbf{A}/\Phi_{0}$ is the
gradient-invariant momentum operator, $\Phi_{0}$ stands for the flux quantum,
$D=lv_{F}/3$ is the diffusion coefficient and $\hbar\omega\equiv\hbar
\omega_{n}=\pi T(2n+1)$ is Matsubara frequency\ ($T$\ is the temperature,
k$_{B}$ = 1).

These equations should be supplemented by formulas for the order parameters
and relation between supercurrent and vector potential. The self-consistency
conditions for the order parameters $\Delta_{S}$ and $\Delta_{A}$ are\
\begin{equation}
\Delta_{S}\ln(T/T_{C0})+2\pi T\sum_{\omega>0}(\Delta_{S}/\hbar\omega-F)=0\\
\end{equation}
\begin{equation}
\Delta_{A}\ln(T/T_{N0})+2\pi T\sum_{\omega>0}(\Delta_{A}/\hbar\omega-\Im)=0
\end{equation}
Here $T_{C0}$ is the superconducting transition temperature when
antiferromagnetic pairing is absent, while $T_{N0}$ is the antiferromagnetic
transition temperature when these is no superconducting pairing. The
self-consistency condition for the vector potential has the usual form (London
electrodynamics):
\begin{equation}
\nabla\times(\nabla\times\mathbf{A)=}\frac{4\pi}{c}j_{S}=i\frac{4\pi}%
{c}eN(0)DT\sum_{\omega>0}(F\mathbf{\Pi}F+h.c\mathbf{.})
\end{equation}
where $N(0)$\ is the density of states on the Fermi surface.\ If an
antiferromagnetic pairing is vanished the Eqs. (2), (3), (5) and (7) restore
those for a nonmagnetic superconductor.

Assuming the two order parameters to be spatially homogeneous, the model (1)
leads to four possible states, each with a specific solutions $(\Delta
_{S},\Delta_{A})$. Namely: (i) the normal state $(0,0)$; (ii) the BCS
superconducting state $(\Delta_{S},0)$; (iii) a purely antiferromagnetic state
$(0,\Delta_{A})$; and (iv) a coexistence phase where both $\Delta_{S}$
and$\ \Delta_{A}$ are non-zero. The phase with the lowest free energy is
stable. The phase diagram for the system can be calculated in terms of the
relative interaction strengths of the two orders (the relationship
$T_{C0}/T_{N0}$) and the fraction of the Fermi surface that is nested (the
results will be published elsewhere.$^{13}$) We will not analyze here all the
phases but only the BCS one when in the sample bulk the antiferromagnetic order is
suppressed. That means that the boundary conditions for Eqs. (2)-(4)
\begin{equation}
G_{0}=\hbar\omega/\sqrt{(\hbar\omega)^{2}+\Delta_{S0}^{2}},F_{0}=\Delta
_{S0}/\sqrt{(\hbar\omega)^{2}+\Delta_{S0}^{2}},\Im_{0}=0
\end{equation}
match the solutions to the Green's function at large distances. Here
$\Delta_{S0}$ is the superconducting order parameter if antiferromagnetic
pairing is absent. 

\section{Basic equations in the vortex geometry}

To study the vortex it is convenient to work in the gauge in which $F$\ and
$\Delta_{S}$ , and $\Im$\ and $\Delta_{A}$\ are real. The coordinate
dependence of all functions is then only through the distance $r$ from the
axis of the vortex and we may define $\varphi\equiv\varphi(\omega
_{n},r)=\varphi_{n}$ and $\theta\equiv\theta(\omega_{n},r)=\theta_{n}$ by such
a way that the Eq.(2) is fulfilled:
\begin{equation}
G(\omega,r)=\cos\varphi\cos\theta,F(\omega,r)=\sin\varphi\cos\theta,\Im
(\omega,r)=\sin\theta
\end{equation}
Due to the cylinder symmetry of a vortex the lines of $\mathbf{A}$ are circles
around it and the relations $\nabla\mathbf{A}=0$, $\mathbf{A}\nabla\varphi=0$
and $\mathbf{A}\nabla\theta=0$ took place. By making use these relations, the
equations of motions for $\varphi$\ and $\theta$\ can be obtained from Eqs.
(3) and (4) in the form:%
\begin{eqnarray}
&&\xi^{2}\{\cos\theta(\nabla^{2}\varphi)-2\sin\theta(\nabla\varphi)(\nabla
\theta)\}= \nonumber \\
&& \frac{1}{2}\xi^{2}\left(  \frac{2\pi\mathbf{A}}{\Phi_{0}}\right)
^{2}\sin2\varphi\cos\theta+\frac{\Delta_{S}(r)}{\Delta_{S0}}\cos\varphi
-\frac{\hbar\omega}{\Delta_{S0}}\sin\varphi
\end{eqnarray}%
\begin{eqnarray}
&&\xi^{2}\{\cos\varphi(\nabla^{2}\theta)+\sin\theta\lbrack\sin\varphi%
\cos\theta(\nabla^{2}\varphi)+\cos\varphi\cos\theta(\nabla\varphi)^{2}%
-2\sin\varphi\sin\theta(\nabla\varphi)(\nabla\theta)]\}\nonumber\\
&&=\frac{\hbar\omega}{\Delta_{S0}}\sin\theta-\frac{\Delta_{A}(r)}{\Delta
_{S0}}\cos\varphi\cos\theta
\end{eqnarray}
where $\xi=(D/2\Delta_{S0})^{1/2}$ stands for the SC correlation length. The
equation for the vector potentials now is took the form
\begin{equation}
\nabla\times(\nabla\times\mathbf{A)=-}\frac{4\pi}{c}Q\mathbf{A},
\end{equation}
where the diamagnetic response function $Q$ reads
\begin{equation}
Q=4\pi e^{2}N(0)DT/\hbar c\times2\pi\sum_{\omega>0}\sin^{2}\varphi\cos
^{2}\theta
\end{equation}
At large distances from the flux axis ( $r\rightarrow\infty$ ), in accordance
with the boundary conditions (8) we have $\varphi\rightarrow\varphi_{\infty
}=arctg(\Delta_{S0}/\omega)$ and $\theta\rightarrow\theta_{\infty}=0$. Then
$Q \rightarrow Q_{0}=c/4\pi\lambda^{2}$ where $\lambda$ is the London
penetration depth is given by the usual dirty limit formula$^{14}$%
\begin{equation}
\lambda^{-2}=\frac{8\pi^{2}}{\hbar c^{2}}e^{2}N(0)D\tanh\frac{\Delta_{S0}}{2T}%
\end{equation}
In the limit of extreme type-II SC the solution to Eqs. (10), (11) and (12)
may be developed by a series of powers of Ginzburg-Landau parameter
$1/\kappa^{2}$ ($\kappa=\lambda/\xi$) . We will calculate the results up
to $O(1/\kappa^{4})$ terms. Note, that for $\kappa\gg1$ and $\xi\leq r$
the expression for vector potential is dominated by the largest term, which,
as follows from Eq. (12), is%
\begin{equation}
A(r)=-\frac{\hbar c}{2e\lambda}K_{1}(r/\lambda).
\end{equation}
Here $K_{1}(x)$ is the imaginary Bessel function of first order. Note that if $\Delta_{A}\equiv0$, $\theta_{n}\equiv0$ our formulas 
restore those considered by Watts-Tobin and Waterworth.$^{15}$

For the purposes of computation, it is convenient to choose $R=r/\xi$ as the
independent variable and to obtain
\begin{equation}
\frac{d^{2}\varphi_{n}}{dR^{2}}+\frac{1}{R}\frac{d\varphi_{n}}{dR}=\frac{1}%
{2}\kappa^{-2}\sin2\varphi_{\infty}K_{1}^{2}(R/\kappa)-\frac{\Delta
_{S}(R)}{\Delta_{S0}}\frac{\cos\varphi_{n}}{\cos\theta_{n}}+\frac{\hbar
\omega_{n}}{\Delta_{S0}}\frac{\sin\varphi_{n}}{\cos\theta_{n}}%
\end{equation}%
\begin{equation}
\frac{d^{2}\theta_{n}}{dR^{2}}+\frac{1}{R}\frac{d\theta_{n}}{dR}%
=\{\frac{\Delta_{S}(R)}{\Delta_{S0}}\sin\varphi_{n}+\frac{\hbar\omega_{n}%
}{\Delta_{S0}}\cos\varphi_{n}\}\sin\theta_{n}-\frac{\Delta_{A}(R)}{\Delta
_{S0}}\cos\theta_{n}%
\end{equation}
Here in the terms of $\propto\kappa^{-2}$ , the zero-order values of
$\mathbf{A(}r\mathbf{)}$, $\varphi(r)$ and $\theta(r)$ are used. Accordingly,
the self-consistency conditions now are determined by%
\begin{equation}
\Delta_{S}(R)=2\pi T\sum_{n\geq0}\sin\varphi_{n}\cos\theta_{n}\{\ln\frac
{T}{T_{C0}}+2\pi T\sum_{n\geq0}1/\hbar\omega_{n}\}^{-1}%
\end{equation}%
\begin{equation}
\Delta_{A}(R)=2\pi T\sum_{n\geq0}\sin\theta_{n}\{\ln\frac{T}{T_{N0}}+2\pi
T\sum_{n\geq0}1/\hbar\omega_{n}\}^{-1},
\end{equation}
and $\Delta_{S0}$ is determined by the relation (18) with $cos\theta_{n}=1$.

We solved these set of equations numerically by an iteration procedure. The
results of our calculations are presented in Figs. 1-3, where the SC and AF order
parameters near the vortex core,\ $\Delta_{S}(R)$ and $\Delta_{A}(R)$ ,
respectively, are shown for different values of the $T_{C0}/T_{N0}$\ ratio,
for a given fraction of the Fermi surface that is nested and at different
temperatures. The solid line represents a coordinate dependence of the
superconducting order parameter $\Delta_{S}(R)$; the dashed line depicts the
antiferromagnetic order parameter $\Delta_{A}(R)$ behavior. As it is seen in
the figures, the size of the "antiferromagnetic" vortex core is on the order of the
superconducting coherence length. I.e., the model predicts that the vortex
core is in fact insulating and antiferromagnetic. Moreover, as $T/T_{C0}$ decreases the region near the flux core where
superconductivity and antiferromagnetism to coexist increases.

\section{Summary}

We investigated the structure of an isolated single vortex for a dirty metal
with competing superconducting (a spin singlet pairing) and antiferromagnetic
(a spin triplet pairing) interactions. We shown that the\ external magnetic
field stabilized the antiferromagnetic order around the superconducting core,
in the region where superconductivity has been suppressed. It is obtained that
the vortex core is insulating and antiferromagnetic, in stark contrast to the
standard normal metal cores of traditional superconductors. The main our
nontrivial result is that as $T/T_{C0}$ decreases the region where
superconductivity and antiferromagnetism to coexist increases.

In recent years, one can see an explosion in the field of vortex dynamics,
especially in the materials of high-$T_{C}$ SC. As is well known, inertial
mass of a fluxion play an important role in vortex dynamics. There are various
mechanisms contributing to the flux inertial mass, and the important one is
due to variation of local electronic density of state in Abrikosov vortices.
The metallic core with states very close to the Fermi energy predicted by
Caroly \textit{et. al}.$^{16}$ is the standard picture for discussion of
inertial mass of the vortex. However, the static, long-ranged, magnetism
associated with the vortex core may have nontrivial influence upon transport
properties of magnetic superconductors in an applied magnetic field and should
be taken into account for systems with competition between superconductivity and antiferromagnetism.

The authors are grateful to V. Chabanenko and N. Hayashi for stimulating discussions.

Fig. 1 Radial coordinate dependence of the vortex superconducting order
parameter $\Delta_{S}(R)$ (solid lines) and antiferromagnetic order parameter
$\Delta_{A}(R)$ (dashed lines) for $\kappa=10$ and different values of
$T/T_{C0}=$ 0.05, 0.5, and 0.8. $T_{C0}=T_{N0}$ ; $\nu\leq1$.  
Fig. 2 Same as in Fig. 1 for $T_{C0}=0.5T_{N0}$, $\nu\leq0.5$.
Fig. 3 Same as in Fig. 1 for $T_{C0}=2T_{N0}$, $\nu\leq1$.


\begin{thebibliography}{99}                                                                                               %


\bibitem[1]{1}A. P. Kampf, Phys. Rep., \textbf{249} , 219 (1994).

\bibitem[2]{2}S.-C. Zhang, Science \textbf{275}, 1089 (1997).

\bibitem[3]{3}D. P. Arovas, A. J. Berlinsky, C. Kallin, and S.-C. Zhang, Phys.
Rev. Lett. \textbf{79}, 2871 (1997); S. Alama, A. J. Berlinsky, L. Bronsard
and T. Giorgi, Phys. Rev. B \textbf{60}, 6901 (1999); H. Bruus, K. A. Eriksen,
M. Hallundb\ae k, and P. Heder\aa rd, B \textbf{59}, 4349 (1999).

\bibitem[4]{4}S. Katano, M. Sato, K. Yamada, \textit{et al.}, Phys. Rev. B
\textbf{62}, R14677 (2000); B. Lake, G. Aeppli, K. N. Clausen, \textit{et
al.,} Science \textbf{291}, 832 (2001); B. Lake, H. M. Ronnow, N. B.
Christensen, \textit{et al.,} Nature \textbf{415}, 299 (2002).

\bibitem[5]{5}Ch. Renner, B. Revaz, K. Kadovski, I. Maggio-Aprile, and \O .
Fischer, Phys. Rev. Lett. \textbf{80}, 3606 (1998).

\bibitem[6]{6}R. Kadono, W. Higemoto, A. Koda, \textit{et al}. Phys. Rev. B
\textbf{69}, 104523 (2004).

\bibitem[7]{7}V. F. Mitrovi\'{c}, E. E. Sigmund, M. Eschring,  \textit{et al.
}Nature \textbf{413}, 501 (2001).

\bibitem[8]{8}J. W. Lynn, S. Skanthakumar, Q. Huang, \textit{et al}. Phys.
Rev. B \textbf{55}, 6584 (1997).

\bibitem[9]{9}Yu. V. Kopaev, Sov. Phys. Usp. \textbf{32}, 1033\ (1989)\ [Usp.
Fiz. Nauk\ \textbf{159}, 567 (1989)].

\bibitem[10]{10}G. G\"{u}ner. Rev. Mod. Phys. \textbf{66}, 1 (1994).

\bibitem[11]{11}V. N. Krivoruchko, JETP \textbf{84, }300 (1997) [Zh. \.{E}ksp.
Teor. Fiz. \textbf{111}, 547(1997)].

\bibitem[12]{12}K.-D. Usadel, Phys.Rev.Lett. \textbf{25}, 560 (1970).

\bibitem[13]{13}V. V. Garkysha and V. N. Krivoruchko, to be published.

\bibitem[14]{14}P. G. de Gennes, \textit{Superconductivity of Metals and
Alloys} (Benjamin, NY, 1966).

\bibitem[15]{15}R. J. Watts-Tobin and G. M. Waterworth, Z.Phys. \textbf{261,}
249 (1973).

\bibitem[16]{16}C.Caroly, P.G. de Gennes, J. Matricon. J. Phys. Lett. (Fr.)
\textbf{9}, 307 (1964).
\end{thebibliography}
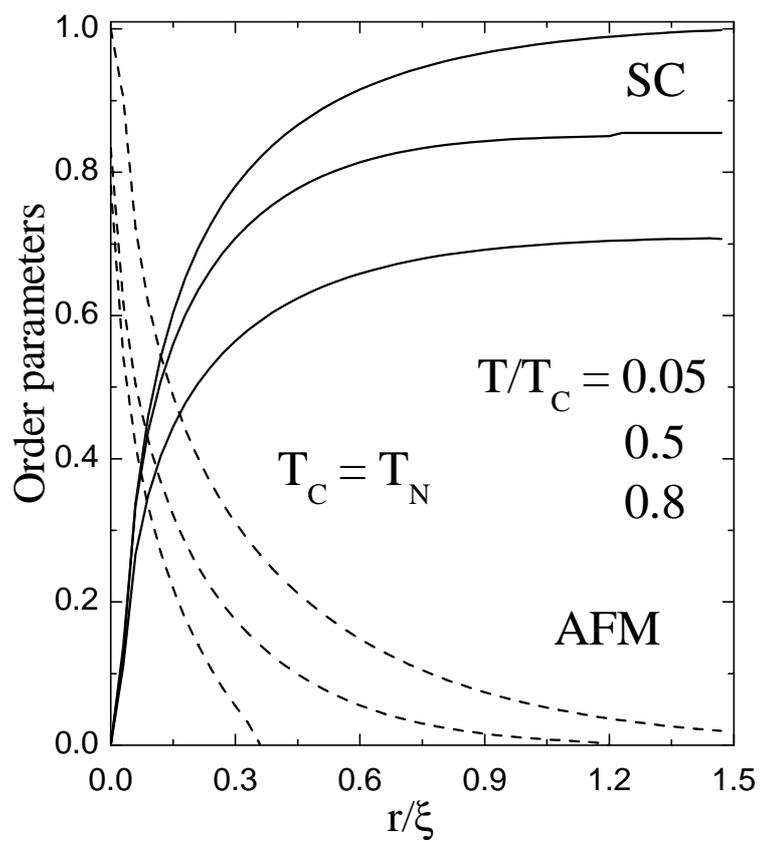
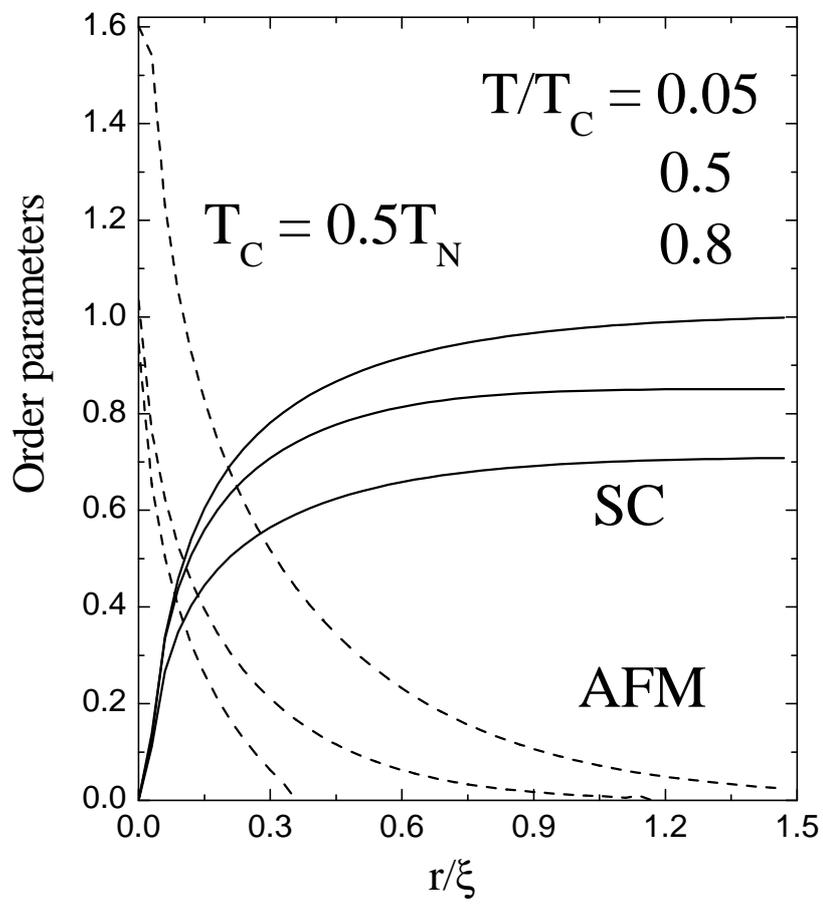
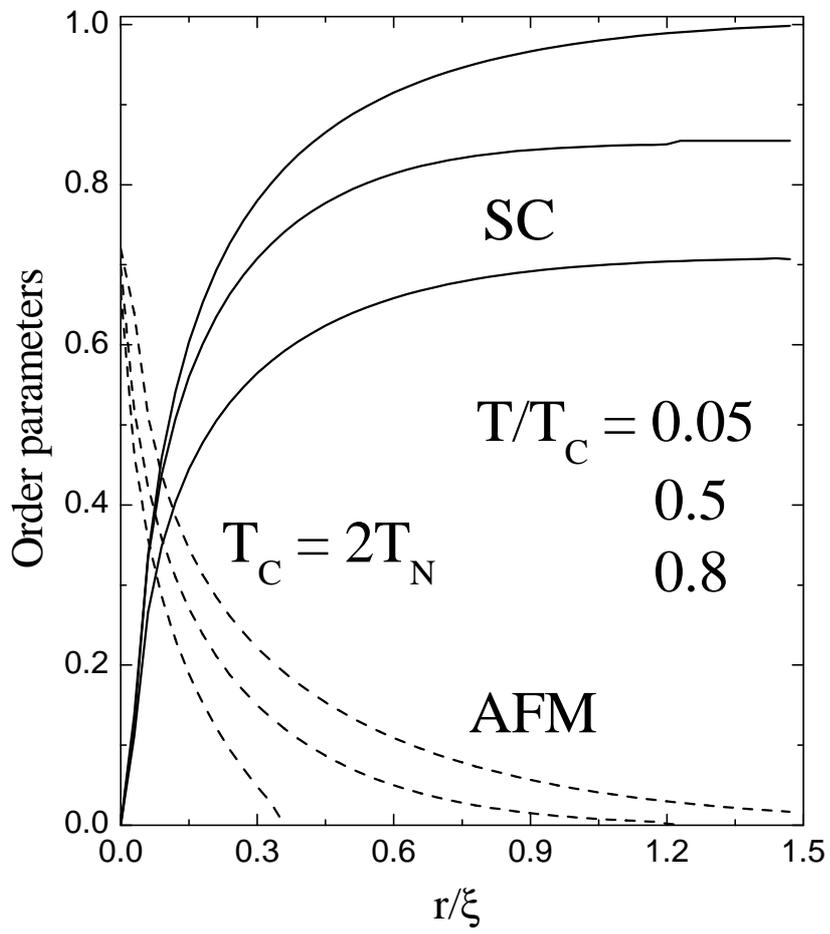
\end{document}